\documentclass[12pt,preprint]{aastex}   

\slugcomment{Submitted to ApJL, 04 September 2003}

\shorttitle{GRB Formation Rate with $E_{p}$--Luminosity Relation}
\shortauthors{Yonetoku et al.}

\begin{document}

\title{GRB Formation Rates inferred from the
 	Spectral Peak Energy -- Luminosity Relation}

\author{D. Yonetoku\altaffilmark{1}, T. Murakami\altaffilmark{1},
T. Nakamura\altaffilmark{2}, R. Yamazaki\altaffilmark{2}, 
A.K. Inoue\altaffilmark{2}, and K. Ioka\altaffilmark{3}}

\email{yonetoku@astro.s.kanazawa-u.ac.jp}


\altaffiltext{1}{Department of Physics, Faculty of Science, 
Kanazawa University, Kakuma, Kanazawa, Ishikawa 920--1192, Japan}
\altaffiltext{2}{Department of Physics, Kyoto University, 
Kyoto 606-8502, Japan}
\altaffiltext{3}{Department of Earth and Space Science, Osaka University,
Toyonaka, Osaka 560-0433, Japan}



\def\d{{\rm d}}
\def\p{\partial}
\def\w{\wedge}
\def\o{\otimes}
\def\f{\frac}
\def\tr{{\rm tr}}
\def\Half{\frac{1}{2}}
\def\half{{\scriptstyle \frac{1}{2}}}
\def\T{\tilde}
\def\RA{\rightarrow}
\def\N{\nonumber}
\def\n{\nabla}
\def\bb{\bibitem}
\def\BE{\begin{equation}}
\def\EE{\end{equation}}
\def\BEA{\begin{eqnarray}}
\def\EEA{\end{eqnarray}}
\def\L{\label}
\def\MG{\mu{\rm G}}
\def\TeV{{\rm TeV}}
\def\EM{E_{\rm max}}
\def\max{{\rm max}}
\def\min{{\rm min}}

\begin{abstract}
We investigate the GRB formation rate based on the relation between
the spectral peak energy ($E_{p}$) and the isotropic luminosity.  
The $E_{p}$--luminosity relation covers the energy range of 
50 -- 2000 keV and the luminosity range of 
$10^{50}$--$10^{54}~{\rm erg~s^{-1}}$, respectively. We find that the 
relation is considerably tighter than similar relations suggested 
previously. Using $E_{p}$--luminosity relation, we estimate the luminosity 
and the redshift of 684 GRBs with the unknown distances and derive 
the GRB formation rate as a function of the redshift. 
For $0 \le z \le 2$ ,  the GRB formation rate is well correlated with 
the star formation rate while it increases monotonously from $z\sim 2$ 
out to $z \sim 12$. This behavior is consistent with the results of 
previous works using the lag--luminosity relation or the 
variability--luminosity relation.

\end{abstract}


\section{Introduction}\label{sec:introduction}

Many ground based telescopes have tried to detect optical afterglows of
GRBs and to measure their redshifts by using the spectral absorption 
and/or emission lines of the interstellar matter in the host galaxy. 
However, the number of GRBs with measured redshift is only a fraction of 
all GRBs detected with BATSE, BeppoSAX, HETE-II and INTEGRAL satellites;  
we have still only about 30 GRBs with the known redshifts. 
The most of them occur at the cosmological distance, and the current record 
holder is GRB~000131 at $z = 4.5$ \citep{andersen}. 
According to the brightness distribution of GRBs with the known redshifts, 
the above satellites should have already detected much more distant GRBs, 
such as at $z \sim 20$ \citep{band03}.  If we can establish a method for 
estimating the intrinsic brightness from the characteristics of the prompt 
gamma-ray emission, we can use the brightness of the GRB as a lighthouse 
to determine the unknown redshifts of majority of GRBs, which enables us 
to explore the early universe out to $z \sim 20$. 

Using the geometrical corrections of the collimated jets, \citet{frail}
and \citet{bloom03} revealed that the bolometric energies released in the 
prompt emission are tightly clustering around the standard energy of
$\sim 1\times 10^{51}~{\rm erg}$. 
Thus, the explosion energy of GRBs can be used as a standard candle as
the supernovae. However, the apparent brightness of GRBs strongly
depends on the jet opening angle and the viewing angle. To use the GRB 
as the standard candle, we need to correct such effects.

Several authors tried to establish a method for estimating the isotropic 
luminosity from the observed GRB properties. Using the lightcurves of 
the prompt gamma-ray emissions, some pioneering works have been done;  
the variability--luminosity relation reported by \citet{fenimore}, 
which indicates that the variable GRBs are much brighter than the smooth 
ones. The spectral time-lag, which is the interval of the peak arrival 
times between two different energy bands, also correlates with the 
isotropic luminosity \citep{norris}. These properties based on the 
time-series data might be due to the effect of the viewing angle from 
the GRB jet \citep[e.g.,][]{ioka, norris02, murakami}.

On the other hand, based on the spectral analyses with the K-correction 
\citep{bloom}, \citet{amati} found the correlation between the 
isotropic-equivalent energy radiated in GRBs and the peak energies $E_{p}$, 
which is the energy at the peak of $\nu F_{\nu}$ spectrum. 
\citet{atteia} suggested a possibility of the empirical redshift indicator, 
which is based on the $E_{p}$ and the arrival number of photons.  

Applying these luminosity indicators to the GRBs with the unknown redshifts, 
their redshifts can be estimated from the apparent gamma-ray brightness. 
As a natural application of the obtained redshift distribution, the GRB 
formation rates are discussed by several authors \citep[e.g.,][]{fenimore,
norris, schaefer01, lloyd, murakami}. Especially, using mathematically
rigid method \citep{efron92, petro93, maloney},
\citet{lloyd} have estimated the GRB formation rates from the 
variability--luminosity relation.  These works give basically the
same results; the GRB formation rate rapidly increases with the redshift 
at $0 \la z \la 2$, and it keeps on rising up to higher redshift 
($z \sim 12$). The GRB formation rate did not decrease with $z$ in 
contrast with the star formation rates (SFRs) measured in UV, optical and 
infrared band \citep[e.g.,][]{madau, bar00, stanway}.

In this paper, we establish a new calibration formula of the redshift based
on the $E_{p}$--luminosity relation of the brightest peak of the prompt
gamma-ray emission from 9 GRBs with the known redshifts. 
Importantly, the uncertainty of our formula is much less than those of the
previous works (lag, variability) as shown in section 3.
Applying the obtained calibration, in section 4, we estimate the
redshifts of 684 GRBs with the unknown redshifts. 
We then demonstrate the GRB formation rate out to $z \sim 12$ and the 
luminosity evolution using the non-parametric method 
\citep{efron92,petro93,maloney,lloyd}.
We emphasize that the GRB formation rate derived by us is based on the 
spectral analysis for the first time, and its uncertainty is well 
controlled in a small level. 
Throughout the paper, we assume the flat-isotropic universe with 
$\Omega_{m}=0.32$, $\Omega_{\Lambda}=0.68$ and 
$H_{0}=72~{\rm{km~s^{-1}Mpc^{-1}}}$ \citep{bennett, spergel}.

\section{Data Analysis}\label{sec:analysis}

We used 9 GRBs (970508, 970828, 971214, 980703, 990123, 990506, 
990510, 991216, and 000131) in the BATSE archive with the known redshifts, 
and focused on the brightest peak in each GRB. We performed the spectral
analysis with the standard data reduction for each GRB. We subtracted
the background spectrum, which was derived from the average spectrum
before and after the GRB in the same data set.

We adopted the spectral model of smoothly broken power-law
\citep{band93}. The model function is described below.
\begin{eqnarray}
N(E) = \left\{ 
\begin{array}{ll}
        A \Bigl( \frac{E}{100~{\rm keV}} \Bigr)^{\alpha}
        \exp(- \frac{E}{E_{0}}), 
        & \\
        A \Bigl( \frac{E}{100~{\rm keV}} \Bigr)^{\beta} \Bigl( 
        \frac{(\alpha - \beta) E_{0}}{100~{\rm keV}}\Bigr)^{\alpha - \beta} 
        \exp(\beta - \alpha), 
        & \\
\end{array}
\right.
\end{eqnarray}
for $E \le (\alpha - \beta) E_{0}$ and $E \ge (\alpha - \beta) E_{0}$,
respectively. Here, $N(E)$ is in units of 
$\rm{photons~cm^{-2}s^{-1}keV^{-1}}$, and $E_{0}$ is the energy at the 
spectral break. $\alpha$ and $\beta$ is the low- and high-energy 
power-law index, respectively. For the case of $\beta < -2$ and 
$\alpha > -2$, the peak energy can be derived as $E_{p} = (2+\alpha) E_{0}$, 
which corresponds to the energy at the maximum flux in $\nu F_{\nu}$ spectra.
The isotropic luminosity can be calculated with the observed flux as 
$L = 4 \pi d_{\rm L}^{2} F_{\gamma}$, where $d_{\rm L}$ and $F_\gamma$ are 
the luminosity distance and the observed energy flux, respectively.

\section{$E_{p}$--Luminosity relation}

\begin{figure}
\plotone{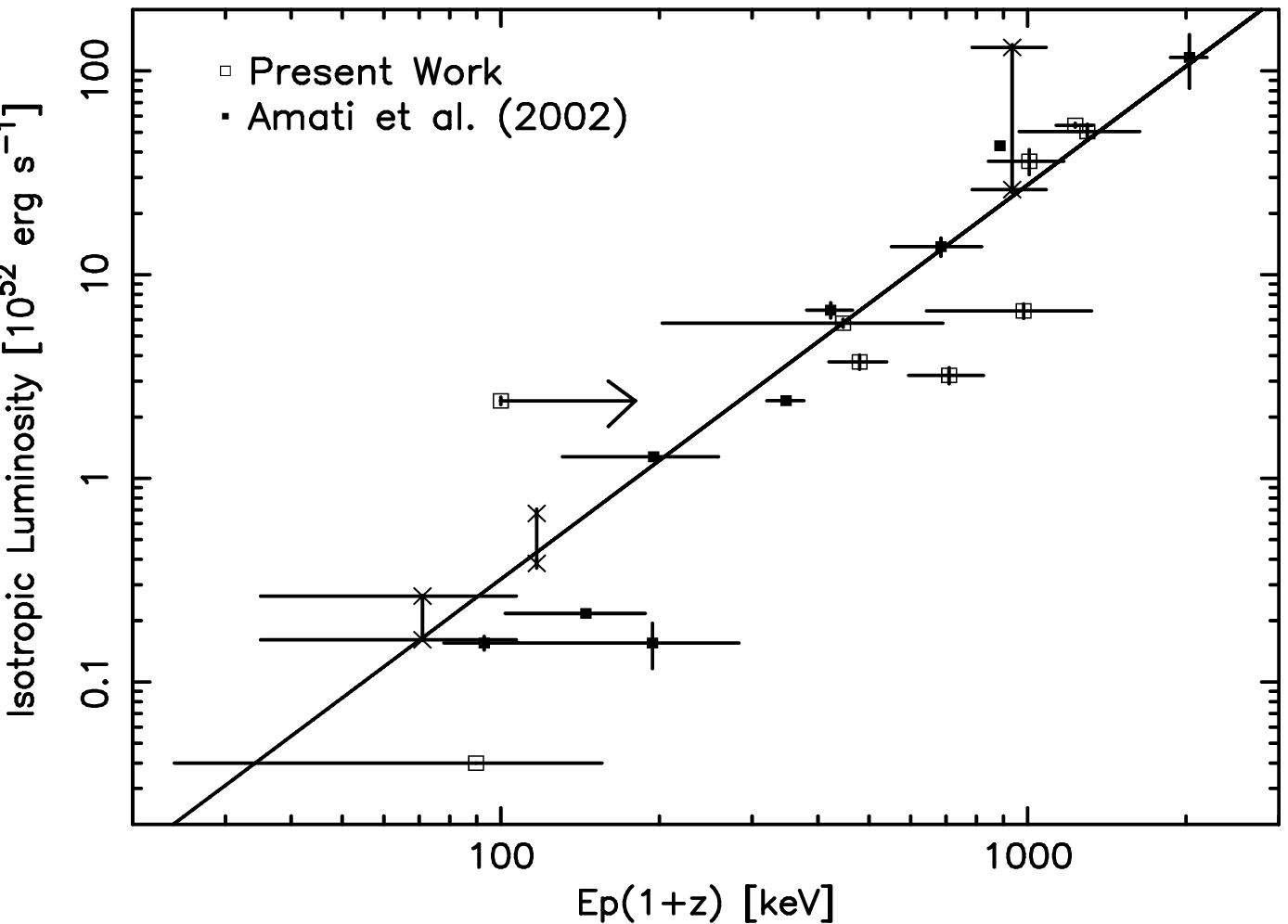}
\caption{The $E_{p}$--luminosity relation. The open squares are our
new results with BATSE. The results of BeppoSAX \citep{amati}, 
which are converted into the energy range of 30 -- 10000 keV,
are also shown as the filled squares and the cross points.
The solid line is the best-fit power-law model for the data. The
linear correlation coefficient is 0.96 for 15 degree of freedom.
\label{fig:ep-lumi1}}
\end{figure}

In figure~\ref{fig:ep-lumi1}, we show the observed isotropic
luminosity in units of $10^{52}~{\rm{erg~s^{-1}}}$ as a function of
the peak energy, $E_{p}(1+z)$, in the rest frame of each GRB. 
For one of our sample (GRB~980703), a lower limit of $E_{p}(1+z)$ is set 
because the spectral index $\beta > -2$.
The results of BeppoSAX reported by \citet{amati} are also plotted in the 
same figure. Here, we converted their peak fluxes into the same energy 
range of 30 -- 10000 keV in our analysis with their spectral parameters.  
There is a good positive correlation between the $E_{p}(1+z)$ and the $L$.  
The linear correlation coefficient including the weighting factors is 
0.957 for 15 degree of freedom (17 samples with firm redshift estimates) 
for the $\log[E_{p}(1+z)]$ and the $\log[L]$.  The chance probability is 
$1.85 \times 10^{-9}$. When we adopt the power-law model as the 
$E_{p}$--luminosity relation, the best-fit function is
\begin{eqnarray}
\label{eq:l-ep}
\frac{L}{10^{52}\ {\rm ergs}} =
(4.29 \pm 0.15) \times 10^{-5} 
\biggl[ \frac{E_{p} (1+z)}{1~{\rm keV}} \biggr]^{1.94 \pm 0.19}
\end{eqnarray}
where the uncertainties are 1~$\sigma$ error. 
This relation agrees well with the standard synchrotron model, 
$L\propto E_{p}^{2}$ \citep[e.g.,][]{zhang, lloyd00}.

\section{Redshift Estimation and GRB Formation Rate}\label{sec:sfr}

The $E_{p}$--luminosity relation derived in the previous section seems to 
be much better estimator of the isotropic luminosity compared with the 
spectral time-lag and the variability of GRBs 
\citep{norris, fenimore, schaefer01} since the chance probability is 
extremely low. In this section, using the $E_{p}$--luminosity relation, 
we try to estimate the isotropic luminosities and the redshifts of the 
BATSE GRBs with the unknown redshifts.

We first selected about 1000 GRBs in a class of the long duration of
$T_{90} > 2~{\rm sec}$ detected by BATSE, and extracted the brightest
pulse in each GRB. We performed the spectral analysis for these peaks with 
the same method described in section~\ref{sec:analysis}. Once we obtained 
observed energy-flux $F_{\gamma}$ and $E_{p}$ at the observer's rest frame, 
we can estimate the redshift with equation~(\ref{eq:l-ep})
and the luminosity distance as a function of the redshift. We could not 
determine $E_{p}$ for $\sim 5$ \% of samples because of $\beta > -2$ 
within 1~$\sigma$ uncertainty, and excluded them. After setting the flux 
limit of $F_{\rm limit} = 1 \times 10^{-7}~{\rm erg~cm^{-2}s^{-1}}$, 
which is based on the dimmest one (GRB~970508), on the data set of $(z, L)$ 
plane, 684 samples are remained. In figure~\ref{fig:z-lumi}, we show 
the sample distribution in $(z, L)$ plane with the truncation by the flux 
limit.

The normalized cumulative luminosity function at each redshift bin 
is also shown in figure~\ref{fig:z-lumi} . The shape of
these luminosity functions are similar to each other, but the
break-luminosity seems to increase toward the higher redshift. This 
indicates that a luminosity evolution is hidden in the $(z, L)$ plane
in figure~\ref{fig:z-lumi}. Therefore, we have to remove the effect of
the luminosity evolution from the data set before discussing the real
luminosity function and the GRB formation rate.
This is because the univariate distribution of the redshift and the 
luminosity can be estimated only when they are independent of each other.

We used the non-parametric method \citep{lynden, efron92, petro93, maloney},
which were used for the Quasar samples and first applied to the GRB samples 
by \citet{lloyd}. The total luminosity function $\Phi(L, z)$ can be written
$\Phi(L,z) = \rho(z) \phi(L/g_{k}(z), \alpha_{s})/ g_{k}(z)$ 
without the loss of generality. Here, each function means the luminosity 
evolution $g_{k}(z)$, the density evolution $\rho(z)$ and the local 
luminosity function $\phi(L/g_{k}(z), \alpha_{s})$, respectively. 
Although the parameter $\alpha_{s}$ represents the shape of the luminosity 
function, we will ignored the effect of this parameter because 
the shape of the luminosity function is approximately same as shown in 
figure~\ref{fig:z-lumi}. We do not mention more about the method, but 
the details are found in \citet{maloney}.

Assuming the functional form of the luminosity evolution as 
$g_{k}(z) = (1+z)^{k}$, which is also used in \citet{maloney} and 
\citet{lloyd}, we convert the data set of $(z, L)$ to $(z, L')$, 
where $L' = L/g_{k}(z)$. We search the best value of $k$ giving the 
independent set of $(z, L')$ within the significance of $1~\sigma$ error, 
and it is found to be $g_{k}(z) = (1+z)^{1.85 \pm 0.08}$. Once we obtain 
the independent data set of the $(z, L')$, we can generate the cumulative 
luminosity function $\psi(L')$ with a simple formula 
\citep[see equation~(14) of][]{maloney}.
In figure~\ref{fig:lumi-func}, we show the cumulative luminosity 
function calculated by the formula. This function is approximately described
by $\psi(L') \propto L'^{-0.3}$ for $0.1 < L'_{52} < 1$ and 
$\psi(L') \propto L'^{-1.2}$ for $1 < L'_{52} < 50$, 
where $L'_{52}=L'/10^{52}\ {\rm ergs}$.

We can also obtain the cumulative number distribution $\psi(z)$
as a function of $z$. The differential form of the function is useful
for the purpose of comparison with the star formation rates in other
wave bands. We convert $\psi(z)$ into the differential form with the
following equation.
\begin{eqnarray}
\rho(z) = \frac{d \psi(z)}{dz} (1+z) 
\left(\frac{dV(z)}{dz}\right)^{-1} \ ,
\end{eqnarray}
where the additional factor of $(1+z)$ comes from the cosmological time 
dilation, and $d V(z)/dz$ is a differential comoving volume.
In figure~\ref{fig:sfr}, we show the relative comoving GRB rate $\rho(z)$ 
in unit proper volume.

\section{Discussion}\label{sec:discussion}

We have investigated the spectral property of the brightest peak of each
GRB with the known redshifts, and have found a fine correlation between the
peak energy $E_{p}(1+z)$ and the isotropic luminosity. While this
correlation against a smaller sample has been pointed out by some authors 
\citep[e.g.,][]{amati, atteia, schaefer03a, schaefer03b}, 
we have succeeded in combining the results of BeppoSAX and BATSE to
describe equation~(\ref{eq:l-ep}). 

\begin{figure}
\plotone{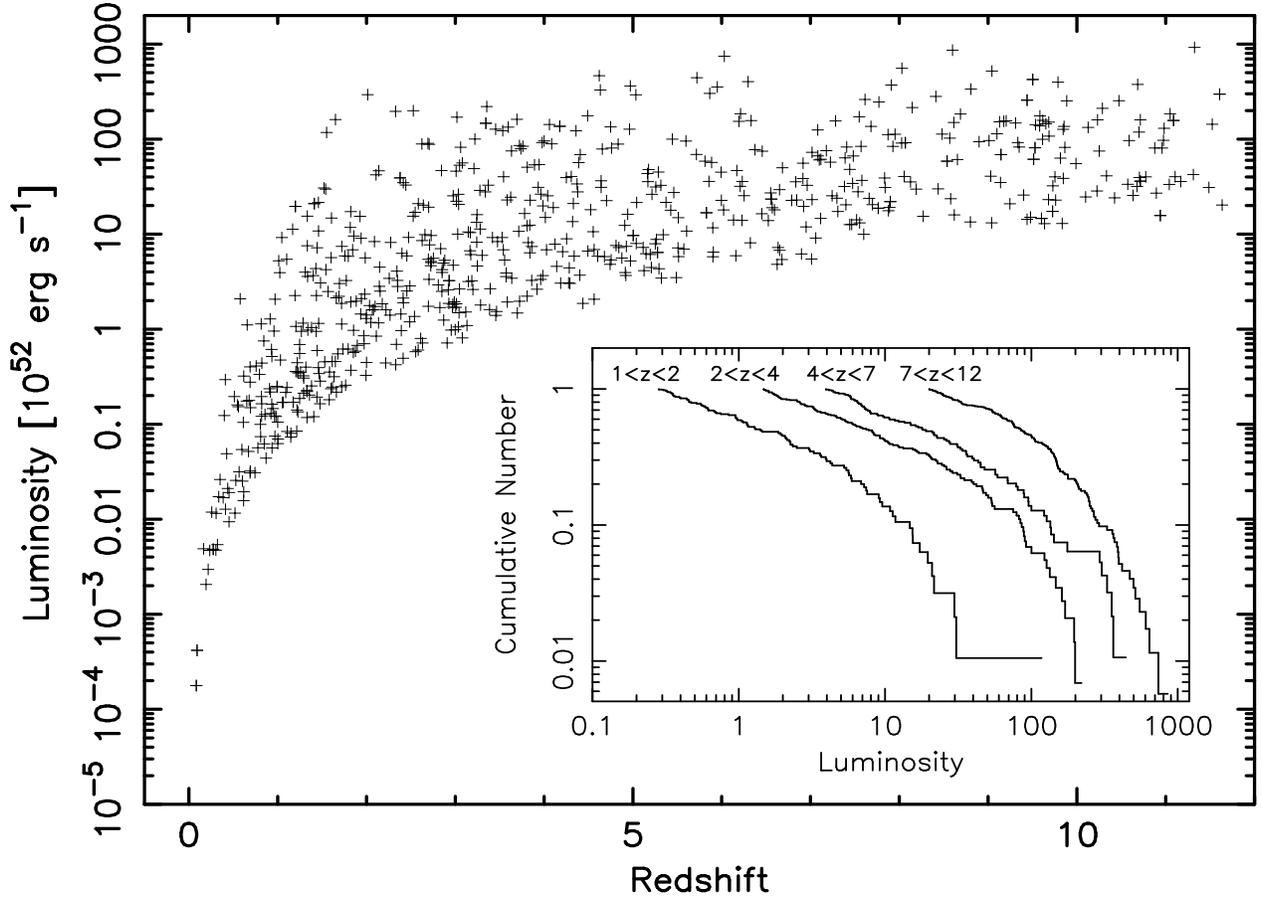}
\caption{The distribution of luminosity vs. redshift derived from
the $E_{p}$--luminosity relation. The truncation of the lower end of
the luminosity is caused by the flux limit of $F_{\rm limit} =
1 \times 10^{-7}~{\rm erg~cm^{-2}s^{-1}}$.  The inserted figure is the
cumulative luminosity function in the several redshift ranges.  
The luminosity evolution exists because the break-luminosity increase
toward the higher redshift.
\label{fig:z-lumi}}
\end{figure}

Using the new $E_{p}$--luminosity relation, we have estimated the redshifts 
of 684 GRBs with the unknown redshifts. We found the 
existence of the luminosity evolution $g_{k}(z) = (1+z)^{1.85 \pm 0.08}$ 
for GRB samples as shown in figure~\ref{fig:z-lumi}. The null-hypothesis 
of the luminosity evolution is rejected about $9 \sigma$ significance. 
\citet{lloyd} also suggest the presence of the luminosity evolution as 
$g_{k}(z) = (1+z)^{1.4 \pm 0.5}$. These two values are consistent with 
each other.  The luminosity evolutions are found in other objects.  
For example, \citet{caditz} and \citet{maloney} estimated the luminosity 
evolution of the QSO samples as $g_{k}(z) = (1+z)^{3}$ and $(1+z)^{2.58}$, 
respectively. Based on the observation of Subaru Deep Field and a 
photometric redshift estimation for $K'$-band selected galaxy samples, 
\citet{kashikawa} found the strong luminosity evolution in the rest UV band 
of their galaxies.

\begin{figure}
\plotone{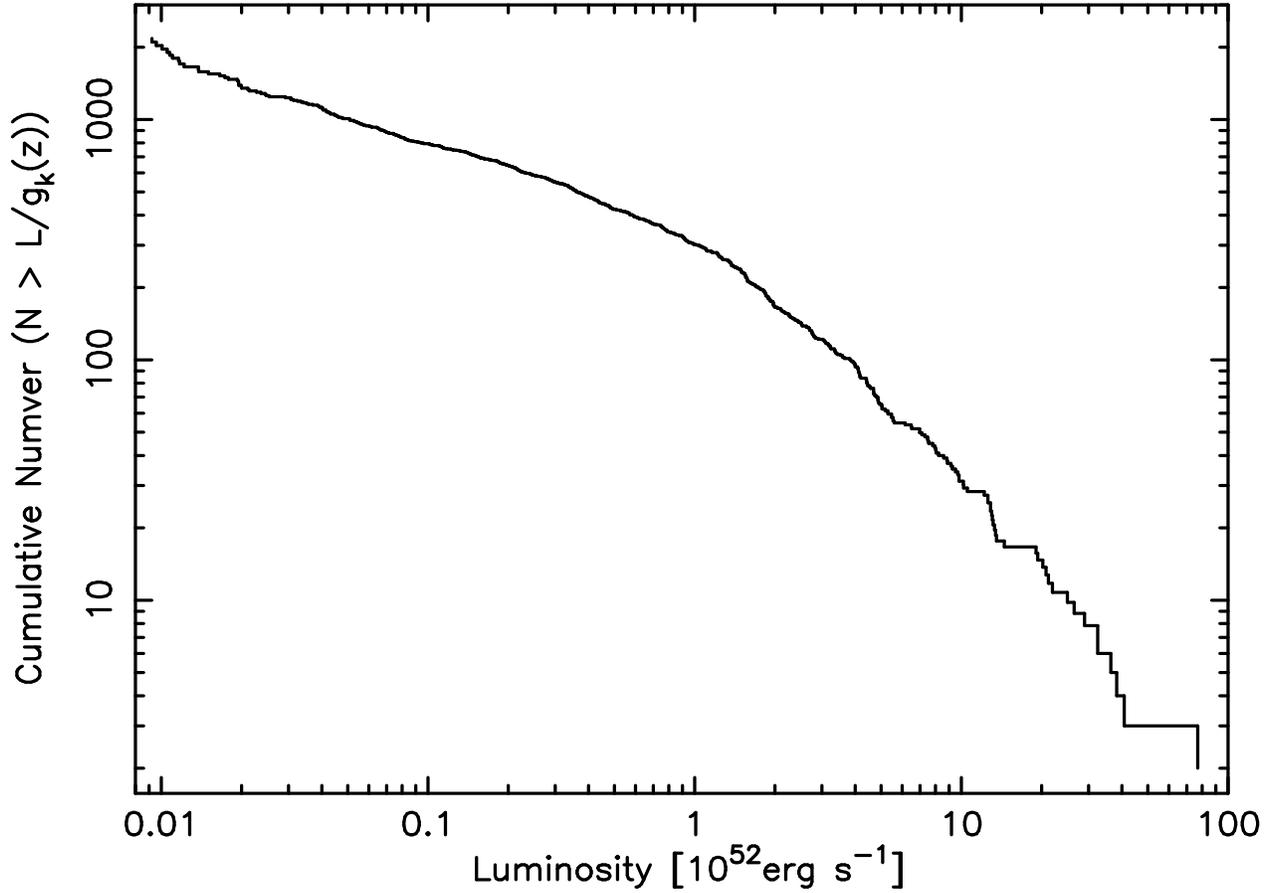}
\caption{The cumulative luminosity function of $L'=L/g_{k}(z)$. 
This function is equivalent to the present distribution at $z = 0$
because the effect of the luminosity evolution $g_{k}(z)$ is removed.
\label{fig:lumi-func}}
\end{figure}

The form of the cumulative luminosity function is independent of $z$
except for the break luminosity, which changes with $z$.
We propose that figure~\ref{fig:lumi-func} might include important
information on the jets responsible for the prompt $\gamma$-ray
emissions and a distribution of their opening angles.
Consider an extremely simple model for a uniform jet with an opening 
half-angle $\theta_j$ and a constant geometrically-corrected luminosity 
$L_0$, which is viewed from an angle of $\theta_v$.
Then, in a crude approximation the luminosity $L$ is given by
\begin{eqnarray}
\label{eq:l-theta}
L = \left\{ 
\begin{array}{ll}
	2 L_{0}\theta_{j}^{-2} \ \ 
        & {\rm for} \ \ \theta_{v} < \theta_{j} \\
	2 L_{0}(\theta_{j}^{6}/\theta_{v}^{8}) \ \ 
        & {\rm for} \ \ \theta_{v} > \theta_{j}. \\
\end{array}
\right.
\end{eqnarray}
For the case of $\theta_{v} > \theta_{j}$, $L$ is proportional 
to $\delta^4$, where 
$\delta=[\gamma(1-\beta\cos\theta_v)]^{-1} \propto \theta_v^{-2}$,
so that the luminosity has the dependence of $\theta_{v}^{-8}$ 
\citep{ioka, yamazaki02, yamazaki}. The dependence of $\theta_{j}^{6}$ 
is determined in order that two functions in equation~(\ref{eq:l-theta}) 
are continuously connected at $\theta_v = \theta_j$.
We also consider the distribution of $\theta_j$ in the form
$f(\theta_j) d\theta_j \propto \theta_j^{-q} d\theta_j$ when
$\theta_{\min}<\theta_j<\theta_{\max}$.
Then, easy calculations show that  in the case of $q < 5/2$, we have
\begin{eqnarray}
\label{eq:2-theta}
N(>L) \propto \left\{ 
\begin{array}{ll}
	L^{-1/4} \ \ 
        & {\rm for} \ L < 2 L_{0} \theta_{\max}^{-2} \\
	L^{(q-3)/2} \ \ 
        & {\rm for} \ 2L_0\theta_{\max}^{-2}<L<2L_0\theta_{\min}^{-2}.
\end{array}
\right.  \ .
\end{eqnarray}
This is a broken power low with the break luminosity 
$2L_0\theta_{\max}^{-2}$. Then if $\theta_{\max}^{-2}L_0 \propto g_{k}(z)$ 
with $q=0.6$, we can roughly reproduce figure~\ref{fig:lumi-func}.
This suggests that either the maximum opening half angle of the jet 
decreases or $L_0$ increases as a function of the redshift. 

This work is the first study to generate the GRB formation rate with
$E_{p}$--luminosity relation. The result indicates that the GRB formation 
rate always increases toward $z \sim 12$. 
This is consistent with the previous works using the GRB variability 
\citep{fenimore, lloyd} and the spectral time-lag
\citep{norris, schaefer01, murakami}. Quantitatively, GRB formation rate
is proportional to $(1+z)^{5-6}$ for $z \la 1$ and to $(1+z)^1$ for 
$z\ga 1$. On the other hand, the SFR measured in UV, optical, and
infrared is proportional to $(1+z)^{2-3}$
for $z \la 1$ \citep[e.g.,][]{lilly,cow99,gla03} and to $(1+z)^{-1-0}$ for 
$z\ga 1$ \citep[e.g.,][]{madau,bar00,stanway,kashikawa}. Therefore, we
find that the ratio of the GRB formation rate to the SFR evolves along
redshift, say $\propto (1+z)^{2-3}$. 

Recently, it has been strongly suggested that the long duration
GRBs arise from the collapse of a massive star 
\citep{galama, hjorth, price, uemura, stanek}.
Hence our result implies that either the formation rate of the massive star 
or the fraction of GRB progenitor in massive stars at the high redshift 
should be significantly  greater than the present value. However, if the 
SFR rapidly increases along redshift as suggested by \cite{lan03}, 
the fraction of GRB progenitors does not change so much.

\begin{figure}
\plotone{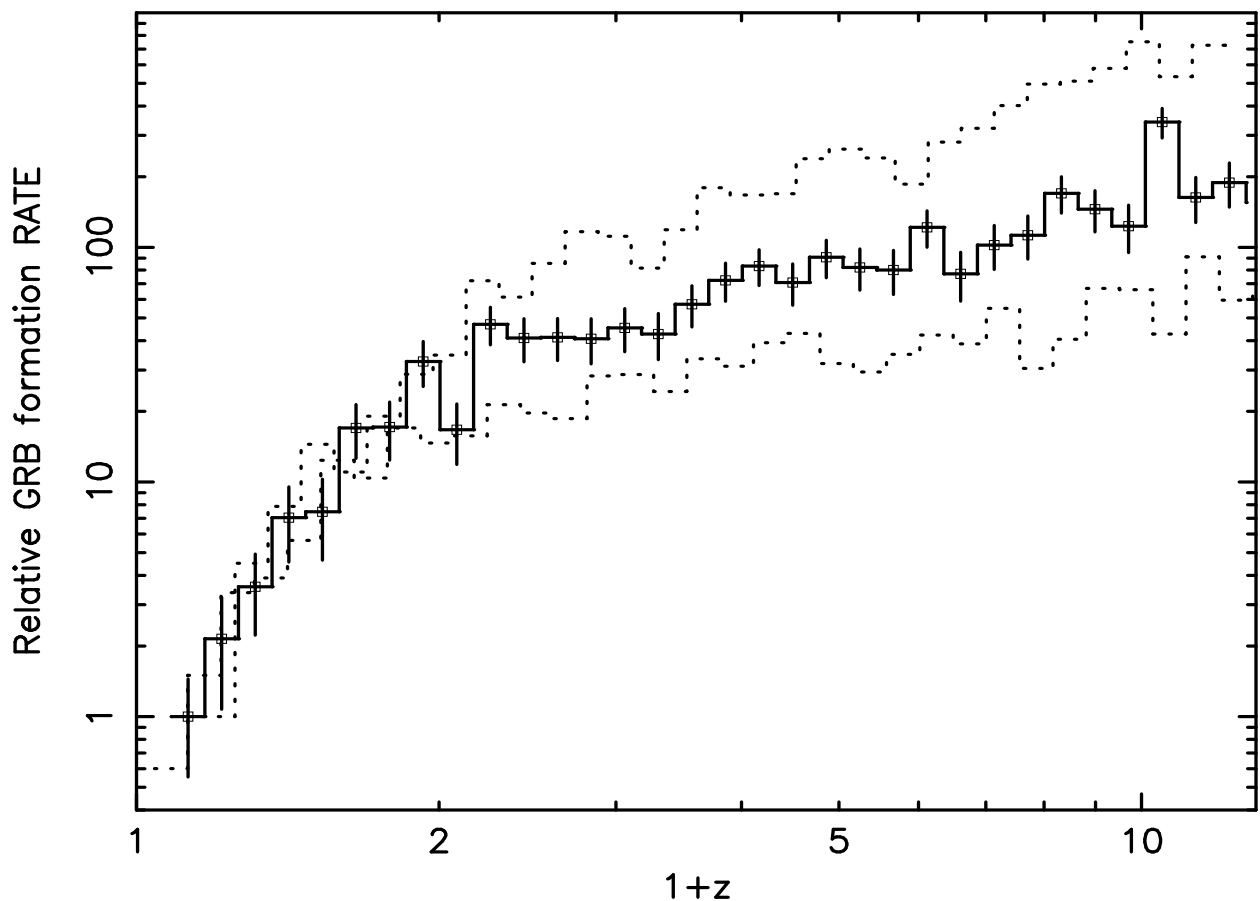}
\caption{The relative GRB formation rate normalized at the first point. 
The solid line is the result based on the best fit of $E_{p}$--luminosity 
relation and two dotted lines indicate the upper and lower bounds 
caused by the uncertainty of $E_{p}$--luminosity relation.
These dotted lines are also normalized and superposed on the best result 
at $0 \le z \le 1$ with the least-square method. The error bars accompanying 
open squares represent the statistical uncertainty of each point.
\label{fig:sfr}}
\end{figure}

The existence of the luminosity evolution of GRBs, 
i.e., $g_{k}(z) = (1+z)^{1.85}$, may suggest the evolution of GRB progenitor 
itself (e.g., mass) or the jet evolution. Although the jet opening angle 
evolution was suggested \citep{lloyd}, in our extremely simple model, 
either the maximum jet opening angle decreases or the jet total energy 
increases. In the former case the GRB formation rate shown in 
figure~\ref{fig:sfr} may be an underestimate since the chance probability 
to observe the high redshift GRB will decrease. 
If so, the evolution of the ratio of the GRB formation rate to the SFR 
becomes more rapid. On the other hand, in the latter case, GRB formation 
rate of figure~\ref{fig:sfr} gives a reasonable estimate.

\acknowledgments

We thank HEASOFT help desk, especially Bryan K. Irby, for the support to
the BATSE analysis tools. This research is supported by Grants-in-Aid from 
the Japanese Ministry of Education, Culture, Sports, Science and 
Technology (15740149, 14204024, 14047212 and 12640302).
R.Y., A.K.I. and K.I. are supported by JSPS Research Fellowship 
for Young Scientists.

\newpage

\end{document}